\documentstyle[prd,aps,epsf,floats]{revtex}
\def\spose#1{\hbox to 0pt{#1\hss}}
\def\lta{\mathrel{\spose{\lower 3pt\hbox{$\mathchar"218$}}
     \raise 2.0pt\hbox{$\mathchar"13C$}}}
\def\gta{\mathrel{\spose{\lower 3pt\hbox{$\mathchar"218$}}
    \raise 2.0pt\hbox{$\mathchar"13E$}}}
 \def\Th{\Theta}                                  
\begin{document}
\preprint{BROWN-HET-...., hep-ph/yymmddd}
\draft

\newcommand{\vp}{\varphi}
\newcommand{\be}{\begin{equation}} \newcommand{\ee}{\end{equation}}
\newcommand{\bea}{\begin{eqnarray}} \newcommand{\eea}{\end{eqnarray}}

\renewcommand{\topfraction}{0.99}
\renewcommand{\bottomfraction}{0.99}
\twocolumn[\hsize\textwidth\columnwidth\hsize\csname 
@twocolumnfalse\endcsname

\title{Thermal Stabilisation of Superconducting Sigma Strings and their 
Drum Vortons}

\author{Brandon Carter \,$^1$ Robert H. Brandenberger \, 
$^{2,3}$  and Anne-Christine Davis \, $^4$}
\bigskip
\address{$^1$
Observatoire de Paris-Meudon, 92 195 Meudon, France;\\
$^2$ Physics Department, Brown University,\\ Providence, R.I. 02912, USA
;\\
$^3$ TH Division, CERN, 1211 Geneva 23, Switzerland (address from
9/15/01 - 3/15/02);\\
$^3$ DAMTP, Center for Mathematical Sciences, University of Cambridge,\\
Wilberforce Road, Cambridge, CB3 0WA, U.K.}

\date{Provisional Draft, 27 December 2001}

\maketitle

%{\bf Abstract}
\begin{abstract}
%We present a new class of defects that can form when there is a small
%symmetry breaking term in the potential. 
We discuss various issues related to stabilized embedded strings in a
thermal background. In particular, we demonstrate that such strings
will generically become superconducting at moderately low
temperatures, thus enhancing their stability.
We then present a new class of defects - drum vortons -
which arise when a small symmetry breaking term is added to the
potential. We display these points within the  
context of the $O(4)$ sigma model, relevant for hadrodynamics below the
QCD scale. This model admits `embedded defects' (topological
defect configurations of a simpler - in this case $O(2)$ symmetric -
model obtained by imposing an embedding constraint) that are
unstable in the full model at zero temperature, but that can be stabilised
(by electromagnetic coupling to photons) in a thermal gas at moderately 
high termperatures. It is shown here that below the embedded defect 
stabilisation threshold, there will still be stabilized cosmic string
defects. However, they will not be of the symmetric embedded vortex type, 
but of an `asymmetric' vortex
type, and are automatically superconducting. In the presence of weak
symmetry breaking terms, such as arise naturally when using the $O(4)$
model for hadrodynamics, the strings become the boundary of a new kind
of cosmic sigma membrane, with tension given by the pion mass.
The string
current would then make it possible for a loop to attain a (classically) 
stable equilibrium state that differs from an ``ordinary'' vorton state 
by the presence of a sigma membrane stretched across it in a drum like 
configuration. Such defects will however be entirely destabilised if 
the symmetry breaking is too strong, as is found to be the case
-- due to the rather large value of the pion mass -- in the hadronic 
application of the O(4) sigma model.

\end{abstract}

]

%\vfill\eject

%\baselineskip 24pt plus 2pt minus 2pt
%\baselineskip 15pt plus 2pt minus 2pt
\vskip 0.4cm

\section{Introduction }
\label{sec:1}

The purpose of this work is to follow up the work of Nagasawa and 
Brandenberger~\cite{Nagasawa:1999iv} who considered the possibility of thermal 
stabilisation, via electromagnetic coupling, of vortex defects, 
i.e. cosmic strings, in a Sigma model characterised by O(4) symmetry 
with a set of degenerate vacuum states having the topology of a 
3-sphere. 

Since the homotopy structure of the 3 sphere is trivial, such a model 
does not have stationary vacuum defects of a topologically stable kind. 
However this model (involving charged and neutral pion fields as well 
as the sigma field) contains a subset of solutions that is identifiable 
as the complete set of solutions of an ``embedded'' model (involving 
just the neutral pion and the sigma field) characterised by O(2) 
symmetry. This embedded model has a set of degenerate vacuum solutions 
having the topology of a circle, and therefore admits stationary vacuum 
vortex defects of a topologically stable kind, which were called
{\it pion strings} in the initial paper \cite{Zhang:1997is} on this subject
\footnote{In this paper we will restrict our attention to the classical
Sigma model and not touch on the rich variety of defects which can
exist when the quantum nature of QCD (in particular at high baryon
density) is taken into account (see e.g.
\cite{Halperin:1998gx,Forbes:2000et,Son:2000fh,Son:2001xd,Kaplan:2001hh,Forbes:2001gj,Balachandran:2001qn} for discussions of such defects).}. 
These stationary 
topological defect configurations of the embedded O(2) model constitute 
what are known
\cite{Vachaspati:dz,Achucarro:1992hs,Barriola:fy,Achucarro:1999it} 
as embedded defects within the framework of 
the full model, but as their energy is not minimised in the broader 
framework of the full O(4) model they will not be stable in this more 
general context. 

The point made by Nagasawa and Brandenberger 
\cite{Nagasawa:1999iv} was that the 
background reference states that are relevant in cosmological contexts 
are commonly not vacuum states but thermal equilibrium states, for which 
topological defects of the embedded O(2) model can be stable as vortex 
defects  of the full model. The possibility of creating such vortex 
defects, i.e. cosmic strings, arises from breaking of the  O(4) symmetry 
by thermal effects mediated by electromagnetic coupling. Such 
stabilisation of an embedded defect (i.e. of a topological defect of the 
embedded O(2) symmetric model) does however require that the product of 
the relevant electric coupling constant $e$ and the temperature $\Th$ 
should be sufficiently large.

The first thing we wish to point out here is that topologically stabilised 
vortex defects of thermal (not vacuum) equilibrium states will exist for 
any non zero value of the product $e\Th$, even if it is very small (as
long as the temperature is higher than the temperature of recombination,
below which the thermal analysis used in this paper breaks down). For large 
values of $e\Th$ these topological defects include the embedded defects 
referred to above. However for smaller values of $e\Th$ the topological 
defects are not configurations of the embedded model, but are of a 
mathematically less trivial kind with the important property that (unlike
their embedded counterparts at higher temperature) they are automatically 
``superconducting'' in the sense of Witten~\cite{Witten:eb}. 
As first observed by Davis and Shellard~\cite{Davis:ij}, such a conductivity 
property allows cosmic string loops to form vortons, i.e. centrifugally 
supported equilibrium states, which under a wide range of conditions will 
actually be stable~\cite{Carter:wu}.

The foregoing considerations are based on the supposition that the 
underlying field model has non thermal vacuum states characterised by 
strict O(4) symmetry, with respect to which the pions are identifiable 
as Goldstone bosons which as such must have zero mass. However for a more 
realistic description, allowing for a finite pion mass that is actually 
observed, the Lagrangian of the model has to be augmented by the inclusion 
of a small intrinsic O(4) symmetry breaking term. This removes the 
degeneracy of the vacuum, as well as of the thermal equilibrium states, 
so there is no longer any possibility of forming a topologically stable 
defect, whether of the vacuum or of a thermal equilibrium state at finite 
temperature.

There is however the possibility at finite temperature of setting up a 
stationary state of a more interesting kind. One of the purposes of 
this article is to consider the construction in such a context of a 
more general kind of (dynamically but not topologically) stable 
equilibrium configuration  that may be described as ``drum vorton'' (or 
``frisbee'')  consisting of a vorton like loop forming the boundary of a 
drum type membrane. 

It is shown that the existence of such stabilised defects is only
possible if the symmetry breaking term is sufficiently small. This
condition may be satisfied in other applications, but it is found
that it does not hold in the case when the O(4) sigma model is applied
in the hadrodynamic context for which it was originally designed.
The failure of the stabilisation mechanism in this particular case is 
attributable to the rather large value of the (destabilising) pion
mass $m_\pi$ in conjunction with the rather small value $e^2 \simeq 1/137$
of the (stabilising) electromagnetic coupling constant.

\section{The bosonic Sigma model }
\label{sec:2}

The following work will be based on the use of a Sigma model
of the usual kind constituting the bosonic sector of the
Schweber Gell-Mann Levy type~\cite{Schwinger:em,Gell-Mann:np} 
hadrodynamic field 
theory as presented in the recent treatise of Walecka~\cite{W95}.
Such a sigma model is given by a Lagrangian density of the form
\be\label{01} {\cal L}= {1\over 16\pi} F^{\nu\mu} F_{\mu\nu}-{1\over 2}
\big(\chi_{|\mu}\,^\ast\!\chi^{|\mu}+\phi_{;\mu}\,^\ast\!\phi^{;\mu} \big) 
- V\, ,\ee
with
\be\label{02} V={\lambda\over 4}\big(\chi\,^\ast\!\chi+ \phi\,^\ast\!\phi
-\eta^2\big)^2 -\varepsilon\,\lambda\, \sigma\, ,\ee
where
\be\label{03} F_{\mu\nu}= A_{\nu;\mu}-A_{\mu;\nu}\, ,\hskip 0.7 cm \chi_{|\mu}=\chi_{;\mu}+ie A_\mu\chi \, ,\ee
in which a semicolon denotes Riemannian covariant differentiation
with respect to the spacetime metric $g_{\mu\nu}$ which we take
to have signature (-,+,+,+), while $^\ast$ denotes complex conjugation. 
In addition to the electromagnetic gauge potential $A_\mu$, the
independent bosonic fields are a set of four real scalar fields
consisting of a pion triplet $\pi_0$, $\pi_1$, $\pi_2$ and a 
singlet, $\sigma=\pi_3$ say, that combine as the complex,
(respectively charge coupled and neutral)  combinations
\be\label{04} \chi=\pi_1+ i\pi_2\, ,\hskip 0.7 cm \phi =\pi_3+i\pi_0
\, ,\ee 
so  that
\be \label{05}\sigma={_1\over^2}(\phi\,+\,^\ast\!\phi)\ee
The other quantities involved are constants, of which $\lambda$ and the usual
charge coupling constant $e$ are dimensionless (and respectively large and 
small, but only moderately, compared with unity) while $\eta$ has the 
dimensions of a mass (with a value about half that of the pion) and
finally $\varepsilon$ has the dimensions of the cube of a mass (that is
small compared with that of the pion).

It is evident that (independently of the local U(1) gauge symmetry
of the electromagnetic part) there will be a global O(4) symmetry
that will act on the sigma pion sector in the limit when the constants
$e$ and $\varepsilon$ are set to zero. This can be seen from the
fact that the corresponding limit for the potential
has the obviously O(4) invariant form
\be \label{06} V\sim {\lambda\over 4}\big( \pi_{0}^{\ 2}+\pi_{1}^{\ 2}+
 \pi_{2}^{\ 2}+\pi_{3}^{\ 2}-\eta^2\big)^2\, ,\ee
(and similarly for the kinetic term)
as $e\rightarrow 0$ and $\varepsilon\rightarrow 0$.

The presence of the symmetry breaking term proportional to $\varepsilon$
is empirically needed~\cite{W95} to account for the finite (observed) value 
of the pion mass $m_\pi$ that is given by the vacuum state value
of the partial derivative
\be\label{107} {\partial V\over\partial (\chi\,^\ast\!\chi)}=
{\lambda\over 2}\big(\chi\,^\ast\!\chi+\phi\,^\ast\!\phi-\eta^2)
\, ,\ee 
The vacuum, i.e the state for which $V$ is minimum, is evidently
characterised by the vanishing of the triplet $\pi_{_0}$, $\pi_{_1}$, 
$\pi_{_2}$ while the value of $\sigma$ will be given by the
larger, $\sigma_{_+}$ say, of the pair, $\sigma_{_+}$, $\sigma_{_-}$
of values where (for vanishing pion fields) $V$ has a local minimum. 
These restricted minima are given by the highest and lowest solutions 
of the cubic equation
\be\label{108} \sigma_{_\pm}(\sigma_{_\pm}^{\, 2}-\eta^2)= 
\varepsilon  \, .\ee
The relevant  values will be given, for small values of $\varepsilon$,
by the expansion
\be\label{109} \sigma_{_\pm}=\pm\eta +{\varepsilon\over2 \eta^2}
\mp{3\varepsilon^2\over 8\eta^5} + {\cal O}\{\varepsilon^3\}\, .\ee
Substituting the vacuum state value $\sigma=\sigma_{_+}$ in the formula
\be m_\pi^{\, 2}={2\partial V\over\partial (\chi\,^\ast\!\chi)}
\ee 
one finds that for low values of $\varepsilon$ the result will 
be given by 
\be\label{110}  m_\pi^{\, 2}= {\varepsilon\lambda\over\eta} +{\cal O}
\{\varepsilon^3\}\, .\ee
  
\section{The (neutral) embedded model }
\label{sec:3}

The configuration space of the preceeding model evidently
includes an embedded subspace characterised by
\be\label{07} \chi=0 \Leftrightarrow \pi_1=\pi_2=0\, ,\ee
that characterises an (electromagnetically decoupled)
``embedded model'' with Lagrangian
\be\label{08} {\cal L}_{\{0\}}= {1\over 16\pi} F^{\nu\mu} F_{\mu\nu}
-{1\over 2}\phi_{;\mu}\,^\ast\!\phi^{\ ;\mu}  
- V_{\{0\}}\ee
for 
\be\label{09} V_{\{0\}}={\lambda\over 4}\big(\phi\, ^\ast\!\phi
-\eta^2\big)^2 -\varepsilon\,\lambda\, \sigma\, ,\ee
whose solutions will all automatically satisfy the 
field equations of the complete model (\ref{01}).

This embedded model does not involve the charge coupling constant 
(whose actual physical value would be given by $e^2\simeq 1/137$) 
but it does involve the other symmetry breaking parameter 
$\varepsilon$. However in the limit when the latter is set to zero
the reduced  model will be subject to a global O(2) symmetry action,
as can be seen from the fact that the corresponding limit for the 
potential has the obviously O(2) invariant form
\be\label{10} V_{\{0\}}\sim {\lambda\over 4}\big( \pi_{0}^{\ 2}+
\pi_{3}^{\ 2}-\eta^2\big)^2\, ,\ee
(and similarly for the kinetic term) for
$\varepsilon\rightarrow 0$.  In this limit there
will be a set of vacuum states with circular topology
characterised by $\phi\,^\ast\!\phi =\eta^2$. One of the
features of this embedded model in the limit of vanishing $\varepsilon$
will therefore be the presence of topological vortex (i.e. string type)
 defects of the vacuum. Such a configuration will be what is 
describable as an embedded defect from the point of view of the
complete theory whose field equations it will also satisfy in the
relevant limit of vanishing $\varepsilon$, but in this broader
framework it will be unstable since there will be no topological
impediment to its decay via the excitation of the
$\chi$ (charged pion) degrees of freedom.

\section{Thermally modified models: general considerations}
\label{sec:4}

Field models such as those described above can be modified so as to 
allow for the effect of a thermal background, with temperature $\Th$, 
by replacing the relevant potential function, ${\cal V}$, by 
an appropriately modified function, ${\cal V}_{_\Th}$. If the independent
field components involved are denoted by $\varphi_i$, for some index
$i$ labelling the relevant degrees of freedom, then one would
expect the effect of small short wavelength fluctuations $\delta\phi_i$
to be approximately describable by an expansion of the form
\be\label{11}  {\cal V}_{_\Th}= {\cal V}
+{\partial {\cal V}\over \partial \varphi_i}\langle\delta\varphi_i\rangle+
{1\over 2}{\partial^2 {\cal V}\over\partial\varphi_i\partial\varphi_j}
\langle\delta\varphi_i\delta\varphi_j\rangle+...\, , \ee
where the angle brackets denote thermal averages. One would expect the 
odd power averages to cancel out, starting with the linear contributions
\be\label{12} \langle \delta\varphi_i\rangle= 0\, , \ee
so the leading contribution will be of quadratic order. One
would expect the short wavelength bosonic fluctuations to behave  
like a simple Bose Einstein radiation gas for which -- using a
formula of Dolan and Jackiw~\cite{Dolan:gu}, for which
a simpler derivation will be provided below in the Appendix -- 
the quadratic contribution will be given simply by
\be\label{13} \langle\delta\varphi_i\delta\varphi_j\rangle
={\Th^2\over 12}\, \delta_{ij}
\, . \ee
Thus under conditions such that (in order for the use of such a thermal
potential to be meaningful at all) the background variation length scale 
is large compared with the thermal length scale -- i.e. the magnitude
of the thermal symmetry 4-vector $\beta^\mu$ with components
$\{\Th^{-1}, 0, 0, 0\}$ in the thermal rest frame --,
but such that the temperature is small enough (i.e. $\beta^\mu$ is
large enough) for the higher order terms in (\ref{11}) to be neglected, 
one is lead to the use of an approximation given by the formula
\be\label{14}  {\cal V}_{_\Th} - {\cal V}=
{\Th^2\over 24}\delta_{ij}{\partial^2 {\cal V}\over
\partial\varphi_i\partial\varphi_j} \, .\ee

\section{Thermally modified sigma models}
\label{sec:5}

In the particular case of the embedded model characterised
by (\ref{08}) it suffices to identify  the components $\varphi$ 
with the real and imaginary parts of the complex vector $\phi$,
and to take ${\cal V}$ to be $V_{\{0\}}$, so that the 
foregoing prescription leads directly to the formula
\be\label{15} V_{\{0\}{_\Th}}-V_{\{0\}} = {\Th^2}{\lambda\over 6}
\big(\phi\,^\ast\!\phi-{1\over 2}\eta^2) \, .\ee
It is to be remarked that this adjustment (in which the final 
term proportional to $\eta$ is dynamically irrelevant because it
does not depend on $\phi$) is such as to preserve the O(2)
symmetry in the limit $\varepsilon\rightarrow 0$: it is
therefore qualitatively uninteresting for moderate values
of $\Th$, though for higher values (above a critical value
$\Th_{\rm c}$) it will have the physically significant
effect of removing the degeneracy (and the consequent spontaneous
symmetry breaking) of the ground state.

The situation is more complicated for the full sigma model
characterised by (\ref{01}), because in addition to the
potential $V$ given by (\ref{02}) there will be another
coupling energy contribution -- coming from the
gauge coupling in the kinetic term -- so that for the purpose of 
applying the prescription (\ref{14}) we need to make
the identification
\be\label{18} {\cal V}=V+{_1\over^2}e^2 \chi\,^\ast\!\chi \, 
A_\mu A^\mu \, .\ee
In view of the gauge invariance, the four components of the 
covector $A_\mu$ should not all be considered to be
dynamically independent. Imposing the thermal gauge
condition $\beta^\mu A_\mu=0$ (which is compatible with the
usual Lorentz condition $A_\mu^{\ ;\mu}=0$) one is left with
three independent components given with respect to the thermal
rest frame by $A_{_1}$, $A_{_2}$, $A_{_3}$. In order to be able
to apply the formula (\ref{13}) on which the prescription
(\ref{14}) is based, the corresponding field components 
$\varphi_i$ must be specified with the appropriate normalisation,
which can be achieved by taking $\varphi_i=A_i/\sqrt{4\pi}$,
($i$= 1,2,3) in order for the (unrationalised) 
kinetic term $F_{\mu\nu}F^{\mu\nu}/16\pi$ to reduce (subject to the
usual Lorentz gauge condition) to   the standard form
${^1/_2}\big(\varphi_{_1;\mu}\varphi_{_1}^{\ ;\mu}+
\varphi_{_2;\mu}\varphi_{_2}^{\ ;\mu}
+\varphi_{_3;\mu}\varphi_{_3}^{\ ;\mu}\big)$. The
corresponding expression for the electromagnetic potential
will therefore be given by 
\be\label{19} {\cal V}-V=2\pi e^2\chi\,^\ast\!\chi\, (
\varphi_{_1}^{\ 2}+\varphi_{_2}^{\ 2}+\varphi_{_3}^{\ 2})\ee

It can thus be seen that the corresponding thermally modified version of
the complete sigma model (\ref{01}) will be given by
\be\label{20} {\cal L}_{_\Th}= {1\over 16\pi} F^{\nu\mu} F_{\mu\nu}-
{1\over 2}\big(\chi_{|\mu}\,^\ast\!\chi^{|\mu} 
+\phi_{;\mu}\,^\ast\!\phi^{;\mu} \big) - V_{_\Th}\, ,\ee
with
\be\label{21} V_{_\Th}=V+{\cal V}_{_\Th}-{\cal V}\, ,\ee
in which, by application of (\ref{14}), the extra thermal contribution 
can be seen to be given by
\begin{eqnarray}\label{22}  V_{_\Th}-V = {{\Th^2} \over 2} e^2\Big({1\over 6}
A_\mu A^{\mu}+ \pi\, \chi\,^\ast\!\chi\Big)\nonumber\\ 
 +{{\Th^2} \over 2}{\lambda\over 2}\big(\chi\,^\ast\!\chi+
\phi\,^\ast\!\phi - {2\over 3}\eta^2) \, .\end{eqnarray}

As in the case of the embedded model, the  group of terms at the end
(i.e. what remains when $e$ is set to zero) is qualitatively uninteresting 
for moderate values of $\Th$, though for higher values (above a critical 
value $\Th_{\rm c}$ that will be evaluated below) it will have the 
physically significant effect of removing the degeneracy (and the consequent 
spontaneous symmetry breaking) of the ground state. 

The part proportional to $e^2$ is more interesting. The first term breaks 
the electromagnetic U(1) gauge invariance, giving an effective
mass $m_\gamma=\Th|e|\sqrt{4\pi/6}$ to the photon. 

The contribution to (\ref{22}) that is of greatest interest for our 
present purpose is the second term, 
$\pi e^2\Th^2 \chi\,^\ast\!\chi / 2$ which will break the O(4) symmetry 
that would otherwise exist in the limit $\varepsilon\rightarrow 0$.

It can be seen from the thermal generalisation 
\be\label{128} {\partial V_{_\Th}\over\partial (\chi\,^\ast\!\chi)}
=\pi{{\Th^2} \over 2} e^2 +{\lambda\over 2}\big(\chi\,^\ast\!\chi
+\phi\,^\ast\!\phi+{{\Th^2} \over 2}-\eta^2) \, ,\ee
of (\ref{107}), that the charged pion field will have an effective mass, 
$m_\chi$ say, that will  be non-vanishing in this thermally modified case, 
even in the limit for which the symmetry breaking parameter $\varepsilon$ 
is set to set to zero, when specified in the usual way by the formula
\be\label{134}  m_\chi^{\ 2}={2\partial V_{_\Th}\over\partial 
(\chi\,^\ast\!\chi)} \, ,\ee
in the thermal equilibrium state where $V_{_\Th}$ is minimised. 

The analogue of the equation (\ref{108}) for the values of $\sigma$
at the restricted minima of $V_{_\Th}$ can be seen to have the form
\be\label{138} \sigma_{_\pm}(\sigma_{_\pm}^{\,2}-\eta_{_\Th}^{\, 2})=
\varepsilon  \, ,\ee
with
\be\label{135} \eta_{_\Th}^{\ 2}=\eta^2-{{\Th^2} \over 2}\, .\ee 
The relevant solution will be given by an 
expansion analogous to (\ref{109}) as
\be\label{139}  \sigma_\pm\simeq\pm\eta_{_\Th} +{\varepsilon\over 2 
\eta_{_\Th}^{\, 2}} \, ,\ee
in the limit when $\varepsilon\ll\eta_{_\Th}^{\, 3}$. 
This inequality will fail to be satisfied near
the critical temperature, i.e. when $\Th\simeq\sqrt{2}\eta$, in which
case the solution to (\ref{138}) will have an order of magnitude
given simply by
\be\label{139a} \sigma_{_\pm}\approx\pm \varepsilon^{1/3}\, .\ee

Substituting the vacuum state value $\sigma=\sigma_{_+}$ in 
(\ref{134}), one obtains a formula of the form
\be\label{140}  m_\chi^{\ 2} = \pi e^2\Th^2+ 
m_{\pi{_\Th}}^{\ 2} \, ,\ee
in which the first term will remain even when $\varepsilon$ is set to
zero. The other term is the square of
the effective mass $m_{\pi{_\Th}}$ of the uncharged pion 
field $\pi_{_0}$, which will be given simply by
\be\label{141}  m_{\pi{_\Th}}^{\ 2} \simeq  
{\varepsilon\lambda\over\eta_{_\Th}} \, ,\ee
which works out as
\be\label{141a}  m_{\pi_\Th}^{\ 2} \simeq m_\pi^{\,2}\Big(1-{{\Th^2} \over
{2 \eta^2}}\Big)^{-1/2} \, ,\ee
in the limit when  $\varepsilon\ll \eta_{_\Th}^{\,3} $, while
for very small values of $\eta_{_\Th}$, i.e. when $\Th\simeq\sqrt{2}\eta$,
it will be given by  
\be\label{141b} m_{\pi{_\Th}}^{\ 2}\approx\lambda\varepsilon^{2/3}
\approx \Big({\lambda \eta^2\over m_\pi^{\,2}}\Big)^{1/3}\, m_\pi^{\, 2}
\, .\ee

\section{Stabilisation of embedded defect}
\label{sec:6}

Nagasawa and Brandenberger pointed out~\cite{Nagasawa:1999iv} that the presence
of an O(4) symmetry breaking term proportional to $\chi\,^\ast\!\chi$
can stabilise the embedded string type defect that arises,
so long as $\Th$ is not too large, as a 
topological defect in the limit $\varepsilon\rightarrow 0$, 
of the embedded  model obtained by setting $\chi=0$ 
in the thermally extended model characterised by (\ref{20}). The
embedded model obtained in this way is given by 
\be\label{24} {\cal L}_{_\Th\{0\}}= {1\over 16\pi} F^{\nu\mu} F_{\mu\nu}-
{1\over 2}\phi_{;\mu}\,^\ast\!\phi^{;\mu}  - V_{_\Th\{0\}}\, ,\ee
with
\be\label{25} V_{_\Th\{0\}}-V_{\{0\}} ={1\over {12}}
\Th^2 e^2 A_\mu A^{\mu}+ \Th^2{\lambda\over 4}
\big(\phi\,^\ast\!\phi-{2\over 3}\eta^2)\, .\ee
Note that this embedded submodel of the thermally extended model
is different from, and more realistic than, the direct  thermal
extension of the original embedded submodel (\ref{08}). It is
to be observed that
$V_{_\Th\{0\}}$ differs from the quantity $V_{\{0\}_\Th}$
given by (\ref{15}) not only by the photon mass term (which could
be got rid of by adopting the more restrictive embedding condition
to the effect that $A_\mu$ should vanish as well as $\chi$, which is
possible because $\chi$ is its only source) but also because,  
unlike $V_{\{0\}_\Th}$, the effective potential $ V_{_\Th\{0\}}$
allows for the effect of thermal excitations of the field $\chi$
even though the embedding condition set its mean value to zero.
This observation serves as a reminder that a fully realistic 
treatment would require the inclusion of further terms allowing for
the thermal excitation of the fermionic degrees of freedom whose
neglect from the outset -- on the grounds that we are considering
cases where their mean value is zero -- was justifiable as a good
approximation for the zero temperature limit,  but less so at finite
temperature.

It can be seen that the potential for this embedded model (\ref{24}) 
will be given by an  expression of the form 
\begin{eqnarray} \label{26} V_{_\Th\{0\}}+\varepsilon\lambda\sigma
& = & {e^2\over {12}}\Th^2 \, A_\mu A^\mu\nonumber \\ &+&
{\lambda\over 4}\big(\phi\,^\ast\!\phi+{{\Th^2} \over 2}-\eta^2\big)^2+C_{_\Th}
\, ,\end{eqnarray}
where $C_{_\Th}$ is a temperature dependent contribution that is constant
in the sense of being independent of the dynamical field variables, and 
that is therefore irrelevant in so far as its effect in the Lagrangian is 
concerned. It is evident that the condition for the degeneracy of the 
ground state and the existence of the embedded vortex defect in the 
limit when $\varepsilon$ is set to zero is that the temperature
should be less than a critical value given simply by $\eta$ i.e.
\be\label{27} \Th < \Th_{\rm c}\, ,\hskip 0.7 cm 
\Th_{\rm c}= \sqrt{2}\eta\, .\ee

The positivity property of the quantity $\chi\,^\ast\!\chi$ whose vanishing 
characterises the embedding under consideration means that the stability
of the embedded solution can be checked simply by verifying the positivity
of the derivative (\ref{128})
on the embedding, $\chi=0$, where it reduces simply to
\be \label{29}
{{\partial V_{_\Th}}\over{\partial (\chi\,^\ast\!\chi)}}_{\chi\,^\ast\!\chi = 0}
=\pi{{\Th^2} \over 2} e^2
+{\lambda\over 2}\big(\phi\,^\ast\!\phi+{{\Th^2} \over 2}-\eta^2)
\, .\ee
In order for this to remain positive even at the core of the defect
where $\phi\,^\ast\!\phi$ goes to zero, it is evidently necessary and
sufficient to have
\be\label{30} \lambda\eta_{_\Th}^{\, 2}<\pi e^2\Th^2\, .\ee
Combining this with the condition (\ref{27}) for the defect to exist
at all, we see that the necessary and sufficient condition for the existence
of stable cosmic strings in the limit when $\varepsilon$ is set to
zero is that the dimensionless ratio $\Th/\eta$ should lie within
the finite range
\be\label{31} \sqrt{2}\Big(1+{2\pi e^2\over\lambda}\Big)^{-1/2}<{\Th\over\eta}
<\sqrt{2}\, ,\ee
a requirement that would be satisfied in a broad range of temperature
if the dimensionless ratio $2\pi e^2/\lambda$ were reasonably large, 
but that is rather restrictive if this ratio is small compared with  
unity as one expects.

\section{The asymmetric vortex defect}
\label{sec:7}

The instability of the embedded $\chi=0$ vortex defect when the temperature
is too low to satisfy (\ref{31}), i.e. when
\be\label{33}  \Big(1+{2\pi e^2\over\lambda}\Big)\Th^2 < 2\eta^2\, ,\ee
does not mean that there cannot be any stable vortex defect, but merely 
that there cannot be one that satisfies the $\chi$ reflection
symmetry condition. The defects that occur in this case must therefore
be of the asymmetric kind that has been recently discussed by
Axenides, Perivolaropoulos, Trodden and Tomaras
\cite{Axenides:1997ja,Axenides:1997sk,Axenides:1998yz}.

For the full sigma model characterised by (\ref{20}) the potential given 
by (\ref{22}) can be seen to be expressible, in a manner analogous 
to (\ref{26}) by 
\begin{eqnarray} \label{34} V_{_{\Th}}+\varepsilon\lambda\sigma
& = &e^2{{\Th^2} \over 2} \Big({1\over 6} A_\mu A^\mu+\pi\chi\,^\ast\!\chi\Big)
\nonumber \\
& + & {\lambda\over 4}\big(\chi\,^\ast\!\chi+\phi\,^\ast\!\phi
-\eta_{_\Th}^{\ 2} \big)^2+C_{_\Th}\, ,\end{eqnarray}
in which $\eta_{_\Th}$ is given by (\ref{135}).

In the limit when $\varepsilon$ vanishes, it is evident that -- if
and only if the inequality (\ref{27}) is satisfied so that
$\eta_{_\Th}^{\, 2}$  is positive -- this model will have a degenerate 
O(2) invariant family of ground states that are the same as those of
the embedded model (\ref{24}), namely the set of states characterised 
by $A_\mu A^\mu=0$ and $\chi\,^\ast\!\chi=0$ but with 
$\phi\,^\ast\!\phi=\eta_{_\Th}^{\, 2}$.

There must therefore exist corresponding topologically stable
vortex defect solutions with a core where $\phi\,^\ast\!\phi=0$.
When the temperature is in the range (\ref{31}) these stable topological
defects will be identifiable with the embedded defects discussed
in the preceding section. However in the lower temperature
range characterised by the inequality (\ref{33}) (which evidently
makes the weaker condition (\ref{27}) redundant) the topologically
stable vortex defects resulting from the potential (\ref{34})
will no longer satisfy the reflection symmetry condition 
$\chi=0$. The solution -- obtained by minimising
the energy -- will presumably be such that $\chi\,^\ast\!\chi$
reaches a finite maximum on the core where $\phi\,^\ast\!\phi$ 
vanishes, with a value that will presumable be comparable with,
but somewhat less than the value for which $V_{_\Th}$ is minimised
subject to the constraint  $\phi=0$, i.e for the value obtained
by solving (\ref{128}) with $\phi\,^\ast\!\phi$ set to zero,
 which gives the value of the upper bound as
\be\label{36} \chi\,^\ast\!\chi=\eta^2-{{\Th^2} \over 2}\Big(
1+{2\pi e^2\over\lambda}\Big)\, ,\ee
a result that evidently has the necessary property of strict
positivity just so long as (\ref{33}) is satisfied.

\begin{figure}
\begin{center}
\leavevmode
\hspace*{-2.1cm}
%\fbox{%\hbox{%
\epsfxsize=7.5cm
\epsffile{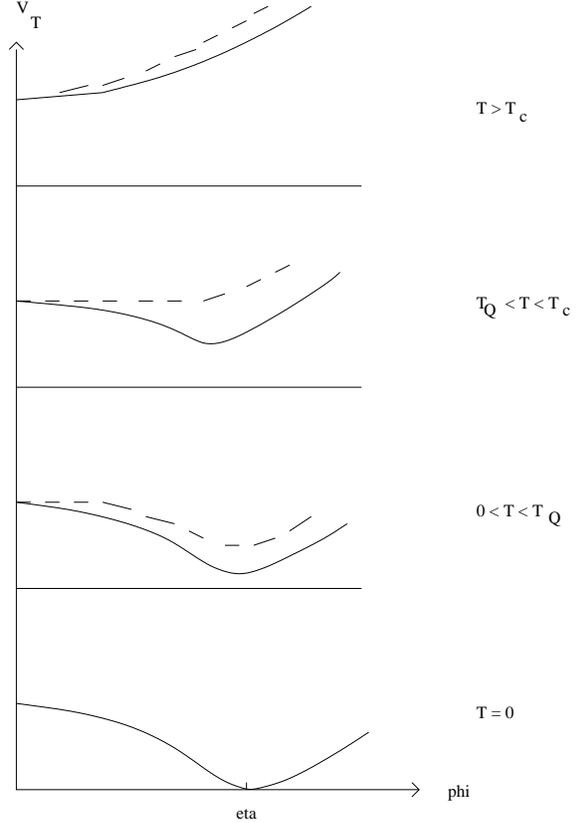}
%}
\end{center}
\caption{Sketch of the effective potential $V_{\Th}$ for
the neutral (solid lines) and charged (broken lines) Higgs 
fields for different temperature ranges. From bottom to
top, the graphs correspond to zero temperature (no stable
string), nonzero temperature below the threshold $\Th_Q$ (when
a stable conducting string with asymmetric core exists),
$\Th_Q < \Th < \Th_c$ (the temperature range for which
the embedded defect with symmetric core is stable), and
$\Th > \Th_c$ (complete symmetry restoration, no string).
In the figure the temperature is denoted by $T$ instead of $\Th$.}
\label{bcfig1}
\end{figure}

\section{Vortex conductivity}
\label{sec:8}

The breakdown of the reflection symmetry property
$\chi=0$ for vortex defects in the low temperature range given by 
(\ref{33}) implies that (unlike the embedded vortex defects
described in Section \ref{sec:6}) the field configuration
in such a vortex will not be uniquely defined but will 
depend on an arbitrary phase angle, $\varphi$ say, defined as
the argument in the expression
\be\label{37}  \chi=|\chi|{\rm e}^{i\varphi}\, .\hskip 0.7 cm\, 
|\chi|=\sqrt{\chi\,^\ast\!\chi}\, .\ee
This means that such a vortex defect will be describable as
a superconducting string in the sense of Witten~\cite{Witten:eb},
meaning that it will be able to carry a current attributable
to slow variation of the phase $\varphi$ along the vortex core.
Its properties will therefore depend on the squared magnitude, 
$\nu^2$ say, of the phase gradient, as specified -- in such a 
way that it will be positive for the case of a spacelike 
gradient with which we shall mainly be concerned here -- by
\be\label{37a} \nu^2=\varphi_{;\mu}\varphi^{;\mu}\, .\ee

In a uniform cylindrically symmetric (Nielsen-Olesen-Kibble type)
configuration described in terms of corresponding cylindrical
time, radial, angular, and longitudinal coordinates,
$\{t,\varpi,\theta, z\}$ of the usual kind, physically well 
defined quantities will be independent of the time $t$ and the
longitudinal coordinate $z$, and will be describable completely
as fields on the 2-dimensional orthogonal space sections
with circular coordinates $\{\varpi,\theta\}$. In a gauge
such that $A_\mu$ has no time or longitudinal components 
the time and longitudinal components of the phase gradient
will be physically well defined and therefore uniform, so the
phase itself can be taken to be given by an expression of
the standard form 
\be\label{38} \varphi=kz-\omega t\, \ee
for some constant angular frequency $\omega$ and wavenumber 
$k$. 

The mechanism described in detail by Peter~\cite{Peter:dw} imposes an
upper limit on the admissible value of the longitudinally Lorentz
invariant combination $\omega^2- k^2$. This limit arises from the
requirement that the charged condensate field $\chi$ should
be effectively confined within a finite length scale, $r_\chi$ say,
whose order of magnitude can be estimated as being given roughly by
\be\label{50} r_\chi\approx {1\over\sqrt
{m_\chi^{\, 2}+k^2-\omega^2}}\, ,\ee
where $m_\chi$ is the relevant mass value for the charged pion field, 
which at zero temperature will be same as that of the uncharged pion 
field, i.e. $m_\chi=m_\pi$ with $m_\pi$ given by (\ref{110}), which 
evidently satisfies $m_\pi^2\geq  \pi e^2\Th^2$, so that there will 
be  a charge dependent term that remains even in the limit of vanishing  
$\varepsilon$ and $m_\pi$. In terms of this mass value $m_\chi$, the 
confinement condition that is needed for $r_\chi$ to remain 
finite will be given simply by
\be\label{51} \omega^2-k^2<m_\chi^{\, 2}\, .\ee   
  
In such a configuration the original 4-dimensional Lagrangian
variational problem reduces to a 2 dimension energy variation
problem for which the original potential function $V_{_\Th}$
has to be replaced by a corresponding 2 dimensional version
\be\label{52} V_{_\Th}^{\,[2]}=V_{_\Th}+{_1\over^2}(k^2-\omega^2) \,
\chi\,^\ast\!\chi \, ,\ee
to allow for the kinetic contributions from  the longitudinal and 
temporal phase variations. Since the ensuing generalisation of 
(\ref{128}) is
\be\label{54} {\partial V_{_\Th}^{\,[2]}\over\partial (\chi\,^\ast\!\chi)}
=\pi{{\Th^2} \over 2} e^2+{_1\over^2}(k^2-\omega^2)+{\lambda\over 2}
\big(\chi\,^\ast\!\chi+\phi\,^\ast\!\phi-\eta_{_\Th}^{\,2})
\, ,\ee
this leads to the replacement of (\ref{36}) by an upper bound that, 
for a spacelike current, $k^2-\omega^2>0$,  is more severe, namely
\be\label{55} \chi\,^\ast\!\chi=\eta^2-{{\Th^2} \over 2}\Big(1+
{2\pi e^2\over\lambda}\Big) -{\nu^2\over\lambda}\, .\ee
in which the Lorentz invariant quantity
\be\label{55a} \nu=\sqrt{k^2-\omega^2} \ee
is identifiable as the phase gradient magnitude that was introduced
in  (\ref{37a}).

The necessity that $\chi\,^\ast\!\chi$  should be positive implies that -- 
just as a timelike current is subject to the ``charge confinement''
limit (\ref{51})  --  at the opposite extreme a spacelike current
will be subject to a ``current quenching''  limit of the kind
originally discussed by Witten~\cite{Witten:eb}, which will be
given in the present application by a relation of the form
\be\label{56} \nu<\nu_{_{\rm Q}}\, ,\ee
where the quenching limit $\nu_{_{\rm Q}}$ is given by the formula
\be\label{56a}  \nu_{_{\rm Q}}^{\,2}=\lambda\, \eta_{_\Th}^{\ 2}- \pi 
e^2\Th^2\, ,\ee
whose right hand side itself satisfies a positivity condition that is 
equivalent to the temperature limit (\ref{33}).
Thus, for fixed temperature $\Th$, then for $\nu < \nu_Q$ the
vortices will be stable. An equivalent way to interpret
this stability analysis is that for fixed current $\nu$, the vortex
will be stable provided $\Th < \Th_\nu$, where $\Th_\nu$ is given by
\be\label{crittemp} 
\Th_\nu^2 = 2 (\eta^2 - {{\nu^2} \over {\lambda}})
(1 + {{2 \pi e^2} \over {\lambda}})^{-1} \, .
\ee 
For $\Th > \Th_\nu$ the current leaks off the vortex.

In the spacelike current case, for which $k^2-\omega^2>0$, there will
be a locally preferred Lorentz frame characterised by $\omega=0$ 
and $k=\nu$ in which the total energy per unit length, $U$ say, will 
exceed the corresponding longitudinal stress magnitude, i.e. the 
string tension $T$ say, by an amount that can be seen from (\ref{52}) to 
be given by \cite{C95}
\be \label{57} U-T=\nu^2\kappa\, ,\ee
where $\kappa$ is the sectional integral of the condensate
amplitude $\chi\,^\ast\!\chi$, i.e.
\be\label{57b} \kappa=2\pi\int \chi\,^\ast\!\chi\, r\,dr\, .\ee
On the basis of (\ref{50}) and (\ref{55}), the order of magnitude of this 
sectional integral can be roughly estimated as
\be\label{57c} \kappa \approx {\lambda \eta_{_\Th}^{\,2}-\pi e^2\Th^2-\nu^2
\over \lambda(m_\chi^{\,2}+\nu^2)}\, .\ee
According to the general principles of conducting string theory~\cite{C95}
the total (sectionally integrated) electromagnetic current density 
$j^\mu$ in the string will have a magnitude given in terms of this
sectional integral by the formula
\be\label{58} j^\mu j_\mu=e^2\mu^2\, ,\hskip 0.7 cm \mu=\kappa\nu \, ,\ee
in which $\mu$ is interpretable as the effective mass per unit phase radian 
winding number, which can be defined in terms of the relevant equation of 
state specifying the energy density $U$ as a function of the angular number 
density $\nu$ by the specification
\be\label{58a} \mu={dU\over d\nu}\, .\ee
It can be estimated using (\ref{57c}) that this effective mass will be 
given roughly by
\be\label{58b} \mu\approx {\nu\over\lambda}\Big( { m_\chi^{\,2} 
+\nu_{_{\rm Q}}^{\,2}\over m_\chi^{\,2} +\nu^2} -1\Big)\, .\ee

\section{The equation of state.}
\label{sec:9}

The value (\ref{55}) of $\chi\,^\ast\!\chi$ is that for which
the derivative vanishes, and thus where $V_{_\Th}^{\,[2]}$ is minimised,
subject to the constraint $\phi\,^\ast\!\phi=0$. The difference
between this minimal value of $V_{_\Th}^{\,[2]}$ and the thermal 
equilibrium state value that is its absolute minimum has a value 
$\Delta V_{_\Th}^{\,[2]}$ that provides a lower bound on the energy density
in the vortex core. This  potential energy density difference will be 
given in the limit of vanishing $\varepsilon$ by the exact formula
\begin{eqnarray}\label{59} \Delta V_{_\Th}^{\,[2]} = 
{\pi e^2 \Th^2\!  +\! \nu^2 \over 4\lambda}&\Big(2\lambda
{\eta_{_\Th}^{\, 2}}\!-\! \nu^2\! -\! \pi e^2\Th^2\Big)
 \, ,\end{eqnarray}
which will be valid so long as it does not exceed the upper limit
\be\label{60} \Delta V_{_\Th}^{\,[2]}\leq {\lambda\over 4}\,
\eta_{_\Th}^{\, 4}\, ,\ee
given by the value of the central energy density in the symmetric 
``embedded vortex'' case that will otherwise be applicable. Thus in 
consequence of (\ref{56}), it can be seen that it will be possible to 
approximate (\ref{59}) by an order of magnitude estimate of the simpler 
form
\be\label{61}  \Delta V_{_\Th}^{\,[2]}\approx 
{\eta_{_\Th}^{\, 2}\over 2}\big(\pi e^2\Th^2 +\nu^2\big)  \, .\ee

This barrier height provides a minimal estimate for the defect energy 
density on the central axis where $\phi\,^\ast\!\phi$ vanishes.
The defect core, meaning the region where $\phi\,^\ast\!\phi$ differs 
significantly from its equilibrium state  value $\eta_{_\Th}^{\, 2}$ 
will be characterised by a radial length scale $r_\phi$ that can be 
estimated from the consideration that energy minimisation will give rise 
to a gradient energy density whose order of magnitude 
$(\eta_{_\Th/} r_\phi)^2$ should be comparable with the 
barrier height $\Delta  V_{_\Th}^{\,[2]}$, i.e. we can expect to
have 
\be\label{62}  r_\phi^{\, 2}\approx {{\eta_{_\Th}^{\, 2}}\over 
{2 \Delta V_{_\Th}^{\,[2]}}}
\, .\ee
In the case of a $\chi$ reflection symmetric ``embedded'' defect that
will apply when the limit  (\ref{60}) is exceeded, this leads
to the simple estimate
\be\label{63} r_\phi\approx {{\sqrt{2}}\over {\eta_{_\Th}\sqrt\lambda}} \, ,\ee
while in the alternative case of an asymmetric defect, i.e. when 
the inequality (\ref{56}) is satisfied, we obtain the estimate
\be\label{64} r_\phi\approx {1\over \sqrt{\pi e^2\Th^2+\nu^2}} \, .\ee

It is to be noticed that in order to obtain confinement to a finite core 
radius it is sufficient but not necessary to have a finite temperature 
$\Th$. Even in the zero temperature limit, for which there is no longer 
any strictly topological stabilisation, there can still in principle
be a confined defect if there is a non vanishing spacelike (but not null 
or timelike) current, i.e. one characterised by a strictly positive value 
of the quantity $\nu^2=k^2-\omega^2$. However in practice a defect that 
depended entirely on this current confining mechanism could not be stable: 
although compatible with the ``quenching'' limit (\ref{56}), the 
necessary current would have to exceed the more stringent upper limit 
imposed, as described below, by the requirement of stability with respect 
to longitudinal perturbations.

It is apparent that in the small $\varepsilon$ limit under consideration 
in the present section, the core radius (\ref{64}) will be of the same 
order of magnitude, $r_\phi\approx r_\chi$, as the charged condensate 
confinement radius given by (\ref{50}). However since the kind of string 
defect we are dealing with is of global rather than local type, its 
energy density will include an unconfined contribution from the gradient 
of the $\phi$ field outside the core where $\phi\,^\ast\!\phi\simeq
\eta_{_\Th}^{\, 2}$. Since this gradient energy density contribution will 
be given approximately by an expression, namely $\eta_{_\Th}^{\,2}/2r^2$, 
whose radial dependence has the inverse square law character that is 
typical of a global vortex defect, its integrated contribution for an 
infinitely long straight string would be logarithmically divergent, so 
that its effective value in practice will be determined by a long range 
cut off radius, $R_\phi$ say, characterising the length scale of macroscopic 
variation of the system. The integrated gradient energy density per unit 
length of string can thus be estimated to have a magnitude of the order of
$\eta_{_\Th}^{\, 2}\, {\rm ln}\{R_\phi/r_\phi\}$. 

On the presumption that, for large $R_\phi$, this gradient contribution will 
dominate, its value in the limit when the current magnitude $\nu$ 
vanishes can be used as a matching condition to fix the constant of 
integration in the solution of (\ref{58a}) using the estimate (\ref{58b}).
We thereby deduce that the equation of state for the string energy 
density $U$ as a function of the angular winding number density $\nu$ 
will be given, as a rough approximation, by an expression of the form 
\be\label{65} U\approx {\eta_{_\Th}^{\,2}\over 2}\,{\rm ln}
\{R_\phi^{\,2}(\nu^2+m_\chi^{\,2})\} -{\nu^2\over 2\lambda}\, ,\ee
in the small $\varepsilon$ limit for which the relevant effective
mass variable will be given by $m_\chi^{\,2}\approx \pi e^2\Th^2$.

On the basis of (\ref{65}),  the corresponding expression for the string 
tension $T$ will be given, according to (\ref{57}) and (\ref{57c}), by
\be\label{66} T\approx {\eta_{_\Th}^{\,2}\over 2}\Big( \,{\rm ln}
\{R_\phi^{\,2}(\nu^2+m_\chi^{\,2})\}- {2\nu^2\over \nu^2
+m_\chi^{\,2} }\Big) +{\nu^2\over 2\lambda} \, .\ee

It is shown in an appendix how an appropriately auxiliary field $\Phi$ 
can be used in a recently formulated procedure~\cite{Carter:1999pq} 
for casting 
such a conducting model into a standard variational form.

\section{The longitudinal stability limit.}
\label{sec:10}

The most physically important quantities derivable from the equation
of state include the extrinsic (wiggle type) perturbation speed
$c_{_{\rm E}}$ and the longitudinal 
(sound type) perturbation speed $c_{_{\rm L}}$, whose squared values
are  given~\cite{C95} by the formulae
\be\label{66a}  c_{_{\rm E}}^{\, 2}= 
{T\over U}\, ,\ee
and (using (\ref{57}), (\ref{58}) and (\ref{58a}))
\be\label{67}  c_{_{\rm L}}^{\, 2}=- 
{dT\over dU}={\nu\, d\mu\over \mu\, d\nu}\, ,\ee
in which the positivity of the right hand sides is a necessary condition
for stability of the short wavength perturbations of the
corresponding (wiggle or longitudinal) kind. 

The largeness of the logarithmic factor ensures the positivity not just
of the energy density $U$ but also of the tension $T$, thus ensuring 
the  satisfaction of the wiggle stability condition that the quantity 
in (\ref{66a}) should be positive throughout the spacelike current 
range $0\leq\nu^2< \nu_{_{\rm Q}}^{\, 2} $ under consideration.  
for longitudinal stability. 

The question of longitudinal stability is less trivial.
As in the analogous case~\cite{Carter:1994hn} of  the local string model 
describing the kind of superconducting vortex 
obtained~\cite{Carter:1994hn}  from the toy bosonic field model originally 
proposed by Witten~\cite{Witten:eb}, the longitudinal stability
requirement,
\be\label{67b} {d\mu\over d\nu}>0\, ,\ee
imposes an upper ``bunching'' stability limit on the physically 
admissible current amplitude that is more severe that the original 
``quenching'' limit (\ref{56}) imposed by the requirement $\mu>0$.
 For the equation of state (\ref{65}), one obtains
\be\label{67c} {d\mu\over d\nu}\approx {(m_\chi^{\,2}+
\nu_{_{\rm Q}}^{\, 2})(m_\chi^{\,2} -\nu^2)\over\lambda
(m_\chi^{\,2}+\nu^2)^2}-{1\over\lambda}\, .\ee
Treating the ratio $m_\chi\nu_{_{\rm Q}}/(m_\chi^{\, 2} 
+3\nu_{_{\rm Q}}^{\, 2})$ as small, which it typically will be 
(it can never exceed 1/3), we see that a necessary and
approximately sufficient condition for satisfaction of the
longitudinal stability limit (\ref{67b}) is given by the
inequality
\be\label{68} {\nu^2\over\nu_{_{\rm Q}}^{\,2}}\lta
{m_\chi^{\,2}\over 3m_\chi^{\, 2}+\nu_{_{\rm Q}}^{\, 2}}\, .\ee
This is interpretable as meaning that the ``bunching'' instability
will set in for a value of the wavenumber $\nu$ that is at
most $1/\sqrt 3$ of the ``quenching'' limit $\nu_{_{\rm Q}}$.

\section{The attached membrane}
\label{sec:11}

Up to this stage our qualitative considerations have been restricted 
to what occurs in the limit $\varepsilon\rightarrow 0$, which should be 
a good approximation in the higher temperature range, above the limit 
(\ref{27}), where the symmetric embedded vortex defect is stable. However 
-- since the pion mass $m_\pi$ given by (\ref{110}) will actually not 
be small compared with $\eta$ but of the same order of magnitude, the 
effects of the symmetry breaking term proportional to $\varepsilon$ will 
be significant in the lower temperature range where the vortex defects 
will be of the asymmetric ``superconducting'' kind.

The effect of this term is to break the degeneracy of the circular set 
of the equilibrium states that were characterised by $\chi\,^\ast\!\chi
=0$, $\phi\,^\ast\!\phi=\eta_{_\Th}^{\, 2} $, by adding unequal adjustment
energy terms, $\delta_{_+}\!V$ and $\delta_{_-}\!V$ say, to the restricted
local minima at $\sigma=\sigma_{_+}$ and $\sigma=\sigma_{_-}\approx
 -\sigma_{_+}$, thereby breaking the degeneracy. On the basis of
(\ref{26}) these adjustments will be given approximately by
\be\label{70} \delta_{_\pm}\!V\simeq -\lambda\varepsilon\sigma_{\pm}\,
\, ,\ee
so that $\delta_{_+}\!V$ is negative whereas $\delta_{_-}\!V$ will be
positive. This means that there will now be an absolute minimum only 
where $\sigma=\sigma_{_+}$. It follows that a circuit round the set
of what were previously degenerate equilibrium states must now cross a 
finite energy barrier whose height, $\delta V_{_\Th}$ say, will be given 
by the difference between the new (absolute) minimum where 
$\sigma=\sigma_{_+}$ and the (restricted) maximum where 
$\sigma=-\sigma{_-}$  (in both cases for 
vanishing values of the pion components) which will be given by
\be\label{69} \delta V_{_\Th}=\delta_{_-}\!V-\delta_{_+}\!V\, .\ee
On the basis of the estimates (\ref{109}) this works out as
\be\label{70a} \delta V_{_\Th}\simeq 2\varepsilon\, \lambda\, \eta_{_\Th}
\, ,\ee
so long as $\eta_{_\Th}^{\, 3}\gg\varepsilon$, while near the critical 
temperature, i.e. when $\Th\simeq\sqrt{2}\eta$, it would have an order of 
magnitude that is obtainable from (\ref{139a}) as
\be\label{70b} \delta V_{_\Th}\approx 2 \varepsilon^{4/3}\lambda \, .\ee

In the case of ordinary cosmic strings it is well known \cite{ShellVil}
that if the symmetry giving rise to the strings is weakly broken
by an explicit symmetry breaking term, then the strings become
boundaries of membrane-like defects. This membrane is the locus
where the phase of the string order parameter changes by $2 \pi$ (if
we consider circling the string in the transverse plane, then the
change in phase is no longer uniform as it would be without symmetry
breaking, but the phase change is localized along one ray in the
transverse plane, i.e. on a membrane in three-dimensional space).

\begin{figure}
\begin{center}
\leavevmode
\hspace*{-2.1cm}
%\fbox{%\hbox{%
\epsfxsize=6.5cm
\epsffile{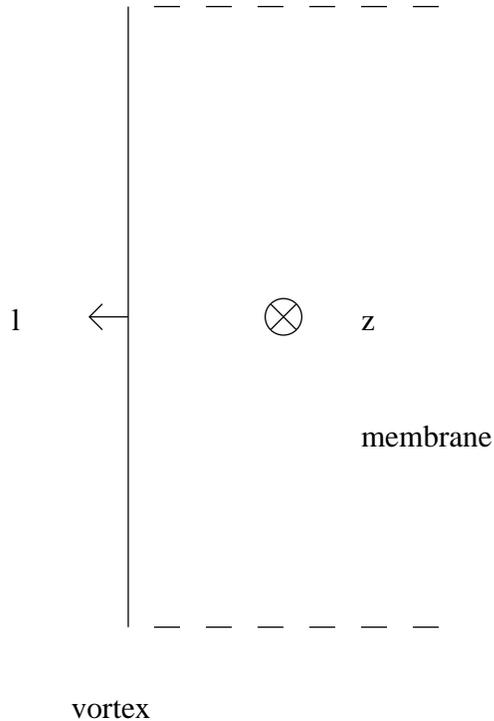}
%}
\end{center}
\caption{Sketch of the vortex and the attached membrane. The
vectors $l$ and $z$ in the figure correspond to the vectors
$\lambda$ and $\zeta$ in the text in Section XII.}
\label{bcfig2}
\end{figure}

The presence of this energy barrier means that it will no longer be 
possible to have a strictly isolated exactly axisymmetric vortex defect, 
but that instead there will be composite defects (of a kind whose
mechanics and classical~\cite{Nagasawa:1994qu} and 
quantum~\cite{Preskill:1992ck}
decay processes have been considered in the context of axion 
theory) that cannot be isolated but must be attached to 
membrane defects. These membrane defects will not actually be ``domain 
walls'' in the strict sense (because the state outside will be the same 
on both sides) but they will have many of the same properties, including a 
thickness length scale $r_\sigma$ say, whose estimation (like that of 
$r_\phi$) can be obtained from the consideration that the energy will be 
minimised when the relevant gradient energy density, of the order of 
$(\sigma_{_+}/ r_\sigma)^2$, is comparable with the relevant barrier 
height $\delta V_{_\Th}$. Thus, as the analogue of (\ref{62}) we obtain 
the general formula
\be\label{71} r_\sigma^{\, 2} \approx {\sigma_{_+}^{\, 2}\over \delta
V_{_\Th}}\, .\ee 

The condition that the symmetry breaking coefficient  $\epsilon$ and hence 
also $\delta V_{_\Th}$ should be small evidently entails that $r_\sigma$
should be correspondingly large. It is only on scales small compared with
this that the phase field distribution outside the string at a 
membrane boundary will retain the axially symmetric distribution
assumed in Section \ref{sec:9}. This radius will therefore act
as the outer cut off introduced in the equation of state function
(\ref{65}). Thus when such symmetry breaking is present it will
normally be appropriate to make the identification
\be R_\phi\approx r_\sigma\, .\ee 

The membrane surface energy density resulting from the concentration
of the phase gradient within the width $r_\sigma$, will be 
identifiable with the tension, ${\cal T }$ say, of the membrane, 
and will be given by the general formula
\be\label{72} {\cal T} \approx r_\sigma\, \delta V_{_\Th}\, .\ee

Although it will  not be topologically stable such a membrane will be 
classically stabilised provided the energy density $\delta V_{_\Th}$
of the membrane forming energy barrier does not exceed the total 
energy density of the string forming energy barrier, as given by 
$\Delta V_{_\Th}-\delta_{_+}\!V$, i.e. by augmenting the original 
string barrier height (\ref{59}) by the adjustment, $-\delta_{_+}\!V$, 
that is  needed to allow for the relative lowering of the ground 
state energy density. Using the analogous expression (\ref{69}) for 
the membrane forming barrier height, this classical stability criterion 
can be seen to be expressible as
\be\label{80} \delta_{_-}\!V <  \Delta V_{_\Th}^{\, [2]}\, .\ee

By rewriting (\ref{59}) in the equivalent form
\be\label{73}  \Delta V_{_\Th}^{\, [2]}={\lambda\over 4}\eta_{_\Th}^{\, 4}
-{(\nu_{_{\rm Q}}^2-\nu^2)^2\over 4\lambda} \, ,\ee
it can be seen that this energy density must always satisfy the
inequality
\be\label{73a} \Delta V_{_\Th}^{\, [2]}<{\lambda\over 4}\eta_{_\Th}^{\, 4}
\, \ee
and hence that (\ref{80}) imposes the requirement
\be\label{74} \eta_{_\Th}^{\,2} >{4\over\lambda}\delta_{_-}V\, .\ee
Evaluating the formula (\ref{70}) for $\delta_{_-}\!V$ using either
of the estimates (\ref{139}) and (\ref{139a}) for $\sigma_{_-}$
leads to the conclusion that classical stability of the membrane
requires that $\Th$ should be low enough for the inequality
\be\label{74a} \eta_{_\Th}^3\geq 4\varepsilon\ee
to be approximately satisfied, with the implication that it is not 
(\ref{139a}) but (\ref{139}) that will be relevant as an order of magnitude 
estimate for $\sigma_\pm$. We thereby obtain the estimate
\be \delta_{_-}\!V \simeq \varepsilon\lambda\eta_{_\Th}\, \ee
which enables us to rewrite the classical stability criterion
(\ref{80}) in the form
\be\label{80a} (\nu_{_{\rm Q}}^{\, 2}-\nu^2)^2\leq \lambda^2\eta_{_\Th}
\big(\eta_{_\Th}^{\, 3}-4\varepsilon\big)\, ,\ee
from which the necessity of  the condition (\ref{74a}) is obvious.
The condition (\ref{74a}) also implies that the barrier height in the 
formula (\ref{71}) will be given generally not by (\ref{70b}) but by 
(\ref{70a}), and hence by (\ref{141}) that the membrane radius will be 
given by
\be\label{76} r_\sigma\approx {1\over m_{\pi_{_\Th}}} \, , \ee 
with $m_{\pi_{_\Th}}$ as given by (\ref{141a}) rather than (\ref{141b}). 
Thus according to (\ref{72}) the membrane tension ${\cal T}$ works out 
to be
\be\label{77} {\cal T}\approx 
 \eta_{_\Th}^{\, 2}\,  m_{\pi{_\Th}} \, ,\ee
or more explicitly
\be\label{77a}  {\cal T}
\approx  \Big(1-{{\Th^2} \over {2 \eta^2}}\Big)^{3/4} \eta^2 \, m_\pi \, .\ee

\section{Drum vorton equilibrium states}
\label{sec:12}

We now examine the way in which the centrifugal effect of the
string current in a closed loop can balance not just the contraction 
of the string tension, as in an ordinary vorton \cite{Davis:ij,Carter:wu,C95}, 
but also the contraction effect of the surface tension of the membrane
that will be stretched across it in a drum type configuration of the 
kind whose investigation has been initiated more recently~\cite{CQ01}.

In so far as the sigma membrane is concerned, the general dynamical 
evolution can be described in terms of a unit vector, $\zeta^\mu$, 
orthogonal to the 3 dimensional sigma membrane world sheet, and on 
its boundary, another unit vector, $\lambda^\mu$, that is tangential
to the world sheet but orthogonal to its boundary. In the two dimensional
string world sheet that constitutes the boundary, there will be a 
preferred orthonormal diad of tangent vectors consisting of a preferred 
timelike unit vector $u^\mu$ and an orthogonal spacelike unit vector 
$\tilde u^\mu$, of which the latter is aligned with the current in the 
spacelike current configuration that we are considering. The set 
$\{ u^\mu, \tilde u^\mu,\lambda^\mu,\zeta^\mu\}$ thus constitutes a 
complete orthonormal tetrad at any point on the bounding string (where 
it is physically well determined modulo sign reversals). In terms of such 
vectors the surface stress energy density tensor, ${\cal T}^{\mu\nu}$,
of the membrane will be given by
\be \label{87} {\cal T}^{\mu\nu}={\cal T}\big (\zeta^\mu\zeta^\nu
-g^{\mu\nu}\big)\, ,\ee
where ${\cal T}$ is the fixed (background temperature dependent) membrane 
tension given by (\ref{77}). The corresponding expression for the surface 
stress energy density tensor, $ T^{\mu\nu}$, of the string at the 
boundary will be given in terms of a variable energy density, $U$, and 
a variable string tension, $T$ by
\be \label{88}  T ^{\mu\nu}= U u^\mu u^\nu- T 
\tilde u^\mu\tilde u^\nu\, .\ee

Systematically using curly script to distinguish quantities associated 
with the 2+1 dimensional world sheet of the membrane from their analogues 
for the 1+1 dimensional world sheet of the boundary string, the relevant 
dynamical equations will be succinctly expressible~\cite{C95} in terms 
of a second fundamental tensor $K_{\!\mu\nu}^{\ \ \rho}$ of the string 
world sheet and of its analogue ${\cal K}_{\!\mu\nu}^{\ \ \rho}$ for the 
membrane. Since the latter evolves freely, its equation of motion will 
be of the simple general form
\be\label{91} {\cal T}^{\mu\nu}{\cal K}_{\!\mu\nu}^{\ \ \rho}=0\, .\ee
The corresponding equation of motion for the string will have the non 
homogeneous form
\be\label{92} T^{\mu\nu}K_{\!\mu\nu}^{\ \ \rho}=f^\rho\, ,\ee
with a force density on the right in which the dominant contribution (at 
least for configurations of large radius) will be produced by the 
attached membrane, whose effect will be given (subject to the 
orientation convention that the membrane tangent vector $\lambda^\mu$ at 
the boundary is outward directed) simply by
\be \label{93} f^\rho={\cal T}^{\rho\nu}\lambda_\nu\, .\ee
In addition to this (in practise inwardly directed) membrane tension
contribution, there will in principle be another (in practise 
outwardly) directed contribution arising from the magnetic field 
produced by the string current. However due to the smallness of the 
electromagnetic coupling constant $e^2\simeq 1/137$ such a ``magnetic 
spring'' effect can be expected~\cite{Peter:pz} to be relatively unimportant.

The simple isotropic form (\ref{87}) of the membrane surface stress 
energy tensor allows us to evaluate the right hand side of (\ref{92}) 
and the left hand side of (\ref{91}) in more explicit form as
\be \label{94} f^\mu =-{\cal T}\lambda^\mu\, ,\ee
and
\be\label{95} {\cal T}^{\mu\nu}{\cal K}_{\mu\nu}^{\,\ \ \rho}
= -{\cal T} {\cal K}^\rho\, \ee
where ${\cal K}^\mu$ is the membrane curvature vector defined by
\be\label{96} {\cal K}^\rho={\cal K}^{\nu\ \rho}_{\ \nu}\, .\ee
Since the world sheet of the membrane is just a hypersurface, with a 
uniquely (subject to a choice of orientation) defined normal 
$\zeta^\mu$, we can work in terms of its second fundamental form 
${\cal K}_{\mu\nu}$ and the trace ${\cal K}$ as defined by
\be {\cal K}_{\mu\nu}={\cal K}_{\mu\nu}^{\,\ \ \rho}\zeta_\rho\, ,
\hskip 0.7 cm {\cal K}={\cal K}_\mu^{\ \mu}={\cal K}^\rho
\zeta_\rho\, ,\ee
thus reducing the generic equation of free motion (\ref{91}) for 
the membrane to the familiar more specialised form
\be\label{97} {\cal K}=0\, .\ee

The membrane dynamical equation (\ref{97}) will of course be trivially 
satisfied in the stationary, flat drum like configurations with which 
we are concerned here. The non trivial  part of the problem is the 
solution of the equation (\ref{92}) that governs the string boundary.  
Specifically our purpose is to look for vorton configurations that are 
characterised as being stationary with respect to a static background 
with respect to a timelike static symmetry generating vector $k^\mu$ 
that not only satisfies the Killing equation
$k_{\nu;\mu}+ k_{;\mu\nu}=0\, ,$ but that is actually covariantly 
constant, 
\be \label{98} k_{\mu;\nu}=0\, .\ee
The stationarity requirement imposes that this Killing vector be tangent 
to the world sheets of the membrane and of its string boundary. If we 
define (modulo another choice of sign) the spacelike unit string tangent 
vector $e^\mu$ to be orthogonal to $k^\mu$ the locally determined stress 
energy eigenvectors will be expressible in the form
\begin{eqnarray}\label{99}  u^\mu= (1-v^2)^{-1/2}\big(k^\mu+ v e^\mu\big) ,
\nonumber\\
 \tilde u^\mu=  (1-v^2)^{-1/2}\big(e^\mu+ v k^\mu\big)\, ,\end{eqnarray}
where $v$ is what will be interpretable as the rotation speed of the 
vorton, which will be given in terms of the phase frequency variables 
introduced in (\ref{38}), as specified with respect to the 
vorton rest frame, by $v=\omega/ k$.

The second fundamental tensor works out in this case to be given by an 
expression of the form
\be \label{100} K_{\mu\nu}^{\ \ \,\rho}=e_\mu e_\nu K^\rho \, ,\ee
in which the curvature trace vector 
\be \label{101} K^\rho= K_{\ \nu}^{\nu\ \rho}\, \ee
will be given simply by
\be \label{102} K^\rho =e^\nu\nabla_\nu e^\rho\, .\ee

For a circular configuration with radius $R$, the curvature vector can 
thus be seen to be given in terms of the radially outward directed unit 
normal $\lambda^\rho$ simply by
\be\label{103}  K^\rho= -{1\over R}\lambda^\rho\, .\ee
By combining this with (\ref{88}) and (\ref{99}) it can be seen that
the left hand side of the string dynamical equation (\ref{92}) will
be given explicitly by
\be\label{104} T^{\mu\nu}K_{\mu\nu}^{\ \ \,\rho} =-{Uv^2-T\over 
(1-v^2) R} \lambda^\rho \, .\ee
Using this in conjunction with the expression (\ref{94}) for the force 
density on the right, which is also proportional to $\lambda^\rho$, the 
dynamical equation (\ref{92}) can be seen to reduce to the simple
explicit form
\be\label{105} U v^2-T=(1-v^2) R\,{\cal T} \, .\ee
In this drum vorton equilibrium equation, it is to be recalled that $U$ 
is the string energy density (its locally preferred rest frame with 
relative motion $v$) and that $T$ is the corresponding (state dependent) 
string tension, while ${\cal T}$ is the (fixed) ``drum'' tension 
characterising the membrane. Thus for an arbitrary string state 
characterised by a chosen energy density $U$ and an associated, 
necessarily smaller,  value of the string tension, $T<U$, it will be 
possible to obtain a circular drum vorton solution with arbitrarily 
large radius $R$ by taking a correspondingly high (but always
subluminal) rotation velocity value given by
\be\label{106} v^2={T+R{\cal T}\over U+R{\cal T}}\, .\ee

\section{Defects in Pion Hadrodynamics}
\label{sec:13}

Since, by (\ref{110}) the symmetry breaking parameter $\varepsilon$
will be given in terms of the observable pion mass $m_\pi$ by
\be\label{150} \varepsilon\simeq {\eta\over\lambda} m_\pi^{\,2}\, ,\ee
the minimal (necessary but not sufficient) defect stability requirement 
(\ref{74a}) can be expressed as the inequality
\be\label{151} m_\pi^{\,2}
\leq {\lambda\eta^2\over 4}\Big(1-{{\Th^2} \over{2 \eta^2}}
\Big)^{3/2} \, .\ee
So long as $\Th$ is not too large compared with $\sqrt{2}\eta$, this will be 
compatible -- albeit rather marginally -- with the observed
ratio, $m_\pi/\eta\approx 2$ due to the fairly 
large value that is empirically~\cite{W95} measured for 
the dimensionless constant $\lambda\approx 25$.

The foregoing requirement is obtained from the condition (\ref{80a})
in the limit for which the phase gradient magnitude $\nu$ has
its maximum ``quenching'' value $\nu_{_{\rm Q}}$ as given by
(\ref{56a}), a value that will in practise be unattainable due
to the ``bunching'' instability limit given by (\ref{68}).

For lesser values of $\nu$,
a necessary -- and, in view of (\ref{74a}), approximately sufficient --
condition for the classical stability condition (\ref{80a}) to be 
satisfied is obtainable, by taking its square root, in the form
\be\label{80b}\label{111} \nu_{_{\rm Q}}^{\, 2}-\nu^2\leq 
\lambda\eta_{_\Th}^{\, 2}-2{\lambda\varepsilon\over\eta_{_\Th} }\, .\ee
Using the expression (\ref{56a}) for the ``quenching'' limit 
 $\nu_{_{\rm Q}}$ and the formula (\ref{141}) for the effective mass
of the uncharged pion, this condition can be rewritten as 
\be\label{152} \nu^2\geq 2\big(m_{\pi_\Th}^{\ 2}-\pi e^2{{\Th^2} \over 2}\big)
\, .\ee

Since the final, charge dependent, term is negative, this requirement
would be satisfied automatically if the symmetry breaking 
term  $m_\pi^{\, 2}$ were small enough to satisfy the condition
\be\label{153} m_{\pi_\Th}^{\ 2}<\pi e^2{{\Th^2} \over 2}\, ,\ee
which can be written more explicitly as
\be\label{153a} 
m_\pi^{\, 2}<\pi e^2\Th^2\sqrt{1-{{\Th^2}\over{2\eta^2}}}\, .\ee
In practise however due to the small value of the electromagnetic
coupling constant $e^2\simeq 1/137$, the relatively large value of the 
symmetry breaking parameter $m_\pi/\eta\approx 2$ ensures that
in the hadrodynamic application the condition (\ref{153}) will fail 
throughout the relevant temperature range $\Th\lta\eta$.

Even when the condition (\ref{153a}) does not hold, the string
defect stability condition (\ref{152}) might still be satisfied
for  sufficiently large value of the current. However as well
as the difficulty of reconciling such a current with the upper
limit on $\nu$ imposed by the bunching stability condition
(\ref{68}), there is the consideration that stability of
the membrane against spontaneous formation of string surrounded
openings requires an energy barrier against formation of 
even the least energetic kind of strings, namely those for
which $\nu$ vanishes.  This suggests that genuinely stable
defect formation will be possible only when (\ref{153a}) is 
satisfied. 

The foregoing reasoning effectively rules out the case of the
hadrodynamic application that motivated this investigation,
but it raises the question of whether the kind of defects
we have been considering might occur in other applications,
involving the same type of O(4) sigma model but with weaker
symmetry breaking. In terms of a dimensionless mass parameter
$\tilde m$ and a dimensionless temperature parameter $\theta$
defined by
\be\label{154} \tilde m={m_\pi\over\eta}\, ,\hskip 0.7 cm 
\theta={{\Th} \over {\sqrt{2}\eta}}\, \ee
the situation may be summed up in the statement that the
defects will be viable if and only if the temperature
is in the limited range for which an inequality of the form
\be\label{155} f\{\theta\}>\tilde m^2 \, \ee
is satisfied, where the function $f\{\theta\}$ simply vanishes
when $\theta$ is greater than unity, and is specified for $\theta<1$
as follows, in a manner that depends on whether $\theta$ is greater
or less that the value $\theta_{_{\rm Q}}$ given by
\be\label{155a}
\theta_{_{\rm Q}}^{\,2}=\Big(1+{2\pi e^2\over\lambda}\Big)^{-1/2}
\simeq 1-{\pi e^2\over\lambda}\, ,\ee
at which the quantity $\nu_{_{\rm Q}}^{\, 2}$
given by (\ref{56a}) vanishes. For the higher range (\ref{30}) one
has
\be\label{165a} \theta\geq \theta_{_{\rm Q}}\ \ \Rightarrow \
f\{\theta\}={\lambda\over 4}\big(1-\theta^2\big)^{3/2}\, ,\ee
by (\ref{151}),
while in the lower range (\ref{33}) one has
\be \label{156b} \theta\leq \theta_{_{\rm Q}} \ 
\Rightarrow \
f\{\theta\}=\pi e^2{{\theta^2} \over 2}\big(1-\theta^2\big)^{1/2}\, ,\ee
by (\ref{153a}). 

Clearly the condition for defect formation will never be satisfied if
${\tilde m}^2$ exceeds the maximum value of $f$, meaning roughly if 
${\tilde m}^2>e^2$. 
In the more interesting case of a cosmological scenario with
\be\label{157} {\tilde m}^2\leq e^2\, ,\ee
the conclusion to be drawn is that as the cosmological temperature 
$\Th$ drops past a first critical value $\Th_{\rm c}$ corresponding
to $\theta=\theta_{\rm c}$ with $\theta_{\rm c}$ given roughly by
$\theta_{\rm c}\simeq 1$, in approximate accordance with (\ref{27}), 
but with a small deviation given in order of magnitude by 
\be \label{158} 1-\theta_{\rm c}^{\, 2}\approx 
\Big({\tilde m^2\over\pi  e^2}\Big)^2\, ,\ee 
the universe would enter a regime in which the condition (\ref{155}) 
is satisfied, so that the defects, in the form of string bounded 
membranes would condense out and  evolve. The strings would be 
superconducting from the outset unless
$\Th_c > \Th_Q$, which implies the (compared to (\ref{157})) 
relatively severe restriction 
\be \tilde m^2\leq {\big(\pi e^2)^{3/2}\over \sqrt\lambda} \ee
is satisfied, and even in this extreme case they would rapidly become 
superconducting as the temperature drops below the value given by 
(\ref{33}) and enters the regime characterised by (\ref{156b}). Due to 
the superconductivity,  some of the defect structure could be 
provisionally preserved (by mechanisms similar to those that have been 
considered for other kinds of string defects~\cite{Brandenberger:1996zp}) 
in the form 
of drum vortons of the kind described in Section \ref{sec:12}. However 
after passing through another lower critical temperature and entering 
a regime characterised roughly by 
\be \theta^2\lta (\pi e^2)^{-1}{\tilde m}^2\, ,\ee  
the condition (\ref{155}) would cease to hold, so there would be 
another phase transition in which any surviving string bounded 
membranes -- including the drum vortons -- would be destroyed. 

\section{Conclusions}
\label{sec:14}
In this paper we have studied the stabilization mechanism for
embedded defects \cite{Nagasawa:1999iv}, with
particular emphasis on the application to the classical
bosonic $O(4)$ sigma model of hadrodynamics \footnote{After
completion of this manuscript a preprint \cite{Ward:2002ci}
appeared which discusses the stabilization of certain
unstable strings and textures by the cosmological expansion.}.

We have seen that below the stabilization threshold for an embedded
defect of the traditional kind (with symmetric core) there will still
be stablized cosmic string defects, but of asymmetric vortex type. These
defects will automatically be superconducting, and this provides them
with an extra stabilization mechanism. These superconducting string
defects are stable above a threshold temperature $\Th_d$ set by the
strength of the explicit symmetry breaking term in the potential, i.e.
by the pion mass in the case of hadrodynamics. In the absence of
explicit symmetry breaking the defects remain stable until the temperature
of recombination, at which point our thermal analysis breaks down.

In the case of explicit symmetry breaking, the superconducting vortices
become boundaries of a new type of membrane-like defects which we call
{\it drum vortons}, across which the change in the phase of the string
order parameter is localized, and whose tension is given
by the symmetry breaking mass, the pion mass in the case of
hadrodynamics.  We have seen that drum vortons can be
stabilized by rotation.

In the case of hadrodynamics, the pion mass is too large for the
superconducing vortices and drum vortons studied here to be stable.
This is due to the large value of the pion mass relative to the
QCD symmetry breaking scale, and due to the large value of the
self coupling constant $\lambda$ relative to the small value of
the gauge coupling $e^2$. However, in many Grand Unified Models,
we expect $\lambda$ to be small, and the explicit symmetry breaking
terms to be absent. In this case, the embedded strings with
asymmetric core studied in this paper and their drum vortons would
be stable.

Thus, we have identified a new class of defects which could be
of great cosmological importance in the early Universe. They
could be used for baryogenesis (see e.g. \cite{Davis:1996zg}) or for the
generation of primordial magnetic fields 
(see e.g. \cite{Brandenberger:1998ew}).
There are also severe cosmological constraints on models which
admit such defects, a topic which we will come back to in a 
subsequent publication \cite{BCD2}.

\centerline{Acknowledgements}

The authors wish to thank J. Blanco-Pillado and A. Zhitnitsky
for discussions. BC and ACD
thank Brown University and RB and ACD thank UBC for hospitality whilst this 
work was in progress. This work was supported in part by the ESF COSLAB 
programme and by a Royal Society-CNRS exchange grant (BC,ACD), by PPARC (ACD),
by the US Department of Energy under Contract DE-FG0291ER40688, Task A (RB), 
and by an Accord between CNRS and Brown University (BC,RB). We are
grateful to Herb Fried for securing this Accord.

\section{Appendix A: mean square amplitude in thermal distribution}
\label{sec:15}

 To derive the coefficient in the ubiquitously useful formula (\ref{13}),
it will suffice to consider the case of a single component field, with
small amplitude $\delta\varphi$ say, which can be analysed as a sum
of contributions from from plane waves with angular frequency $\omega$
in different directions.  From any such plane wave wave contribution, 
the mean square field fluctuation amplitude will receive an
infinitesimal  contribution  $d\langle\,(\delta\varphi)^2\rangle$
that will be related to the corresponding infinitesimal contribution
$d{\cal E}$ to the energy density ${\cal E}$ by a proportionality formula 
that (subject to use of the standard field normalisation convention as 
above) will have the simple form
\be\label{A1} d{\cal E} =\omega^2\, d \langle\,(\delta\varphi)^2
\rangle \, .\ee

In a thermal distribution with temperature $\Th$, the energy density
contribution corresponding to an infinitesimal angular frequency 
range $d\omega$ will be given (in our units, for which the speed of 
light $c$, the Boltzmann constant $k$, and the Dirac Planck constant
$\hbar=h/2\pi$ are all set to unity, i.e. $c=k=\hbar=1$) by the
well known Bose-Einstein gas formula
\be\label{A2}
 d{\cal E}={\omega^3 d\omega\over 2\pi^2\big({\rm e\,}^{\omega/\Th}-1
\big)}\, .\ee
Combining this with (\ref{A1}), and integrating with the substitution
$u=\omega/\Th$,  we find that the total mean square fluctuation amplitude 
will be given by
\be \langle\,(\delta\varphi)^2\rangle={\Th^2\over 2\pi^2}
\int_{_0}^{_\infty} {u\, du\over {\rm e }^{\, u}-1} \, .\ee
Since the integral involved is well known to be given as a Riemann 
zeta function by
\be \int_{_0}^{_\infty} {u\, du\over {\rm e }^{\, u}-1} =\zeta\{2\}=
{\pi^2\over 6} \, ,\ee
we immediately obtain the simple final formula
\be \langle\,(\delta\varphi)^2\rangle={\Th^2\over 12}\, ,\ee
of which the required multicomponent generalisation (\ref{13})
is now an obvious corollary.

\section{Appendix B: Variational formulation}
\label{sec:16}

The conducting string model set up in Sections \ref{sec:8} and
\ref{sec:9} can easily be cast into variational form in terms of an 
action integral 
\be\label{B1} {\cal I} =\int\, {\cal L}\, |\gamma|^{1/2}\,
 d^2\sigma\, ,\ee
that is taken over a the string world sheet with internal coordinates 
$\sigma^a$ ($a=0,1$) and corresponding induced metric $\gamma_{ab}$
for a suitably chosen Lagrangian density scalar ${\cal L}$.
In particular, in terms of the phase scalar $\varphi$ used above
and of an appropriately specified auxiliary scalar $\Phi$,
this Lagrangian can be given the standard form
\be\label{B2} {\cal L} =-{1\over 2}\Phi^2\, \varphi_{|a}\varphi^{|a}
- V\{\Phi\}\, ,\ee
with $\varphi$ and $\Phi$ as independently variable 2-surface
supported fields using the general prescription~\cite{Carter:1999pq} 
\be\label{B3} V={1\over 2}\big(U+T)\, \ee
with 
\be\label{B4}  \varphi_{|a}\varphi^{|a}=\nu^2\, ,\hskip 0.7 cm
\kappa=\Phi^2\, .\ee

In the present application to the model characterised by the equation 
of state (\ref{65}), this prescription simply gives
\be\label{B5} \Phi^2={\nu_{_{\rm Q}}^{\,2} -\nu^2\over
\lambda\,(m_\chi^{\,2}+\nu^2)}\, ,\ee
and hence
\be\label{B6}  \nu^2={\nu_{_{\rm Q}}^2-m_\chi^{\, 2}\lambda\,\Phi^2
\over 1+\lambda\,\Phi^2}\, ,\ee
so it immediately follows from (\ref{65}) and (\ref{66}) that the 
required potential function $V\{\Phi\}$ will be given by
\be V={m_\chi^{\,2}\lambda\,\Phi^2-\nu_{_{\rm Q}}^{\,2}\over
m_\chi^{\,2}+\nu_{_{\rm Q}}^{\,2}}+
{1\over 2}\,\ln\Big\{ {R_\phi^{\,2}(m_\chi^{\,2}+
\nu_{_{\rm Q}}^{\,2})\over 1+\lambda\,\Phi^2}\Big\}\, .\ee

\end{document}